\pdfoutput=1 
\documentclass[11pt]{article}
\usepackage{amsfonts}
\newcommand{\algname}{SeCluD}
\usepackage{subfigure}
\usepackage{fullpage}
\usepackage{url}
\usepackage{color}
\usepackage{epic}
\usepackage{eepic}
\usepackage{amsmath}
\usepackage{graphicx}
\usepackage{listings}
\usepackage{color}
\usepackage{tikz}
\usetikzlibrary{positioning}
\usetikzlibrary{arrows}
\usetikzlibrary{shapes}
\usetikzlibrary{decorations.pathreplacing}
\usetikzlibrary{calc}
\usepackage{graphicx}
\usepackage{datetime}
\usepackage{booktabs}
\usepackage{placeins}
\usepackage{enumitem}
\usepackage{siunitx}
\usepackage{nth}
\newcommand{\set}[1]{\left\{ #1\right\}}

\newcommand{\binomial}[2]{\binom{#1}{#2}}

\newcommand{\realrange}[2]{\left[#1, #2\right]}

\newcommand{\unitrange}[2]{\realrange{0}{1}}

\newcommand{\prob}[1]{{\mathbf{P}}\left[#1\right]}

\newcommand{\expect}{{\mathbf{E}}}

\newcommand{\Oh}[1]{\mathcal{O}\!\left( #1\right)}

\newcommand{\llabel}[1]{\label{\labelprefix:#1}}
\newcommand{\labelprefix}{} 

\newcommand{\discussionsize}{\small}

\newcommand{\punkt}{\enspace .}

\newenvironment{code}{\noindent\normalsize
\begin{tabbing}%
\hspace{2em}\=\hspace{2em}\=\hspace{2em}\=\hspace{2em}\=\hspace{2em}\=%
\hspace{2em}\=\hspace{2em}\=\hspace{2em}\=\hspace{2em}\=\hspace{2em}\=%
\kill}{\end{tabbing}}

\newcommand{\labelcommand}{}
\newcommand{\captiontext}{}
\newsavebox{\codeparam}
\newcounter{lineNumber}
\newenvironment{disscodepos}[3]{%
\renewcommand{\labelcommand}{#2}%
\renewcommand{\captiontext}{#3}%
\sbox{\codeparam}{\parbox{\textwidth}{#3}}%
\begin{figure}[#1]\begin{center}\begin{code}\setcounter{lineNumber}{1}}{%
\end{code}\end{center}\caption{\llabel{\labelcommand}\captiontext}\end{figure}}

{\end{disscodepos}}

\newcommand{\Is}{\mbox{\rm := }}

\newdimen\endofsize\endofsize=0.5em
\def\endofbeweis{~\quad\hglue\hsize minus\hsize
                 \hbox{\vrule height \endofsize width
\endofsize}\par}

\newcommand{\score}[2]{\delta_{#1}^+(#2)}
\newcommand{\scale}{\epsilon}

\begin{document}
\title{Faster Exact Search using Document Clustering}
\author{Jonathan Dimond and Peter Sanders\\
 Karlsruhe Institute of Technology, Karlsruhe, Germany\\
\url{sanders@kit.edu,mail@dimond.de}}

\maketitle
\thispagestyle{empty}

\begin{abstract}
  We show how full-text search based on inverted indices can be accelerated by
  clustering the documents without losing results (\algname\ -- {\bf Se}arch
  with {\bf Clu}stered {\bf D}ocuments). We develop a fast multilevel clustering
  algorithm that explicitly uses query cost for conjunctive queries as an
  objective function. Depending on the inputs we 
  get up to four times faster than non-clustered search. The resulting clusters are also
  useful for data compression and for distributing the work over many machines.
\end{abstract}
\setcounter{page}{0}
\section{Introduction}
\label{s:intro}

Full-text search is one of the enabling techniques of the information society
since it is needed for all kinds of search engines. The approach most used in
practice uses \emph{inverted indices}: Consider a set $D$ of $n$
\emph{documents}. Each document contains a set of \emph{terms} from a dictionary
$T$. The index stores for each term $t\in T$ a \emph{posting list} of document
IDs where it occurs. A typical query asks for documents containing a set of two
or more terms. This query can then be answered by intersecting the corresponding
posting lists. Unfortunately, for large inputs, the lists become huge incurring
substantial energy consumption for full-text search. For example, for web
search, thousands of machines can be involved in answering a single query.  What
helps is that two sorted sequences of document IDs can be intersected in time
linear in the size of the \emph{smaller} list \cite{SanTra07}.

The starting point for this paper was the observation that we can boost the
impact of such advanced set intersection algorithms by distributing the
documents to clusters where terms are distributed as nonuniformly as possible.
Lets consider a simple example to illustrate this effect. Suppose we want to
know all documents that contain both term $a$ and term $b$. Assume we have four
clusters with the following number of occurrences of $a$ and $b$.
\begin{center}
    \begin{tabular}{crr|r}
        \textbf{Cluster} & {\#$a$} & {\#$b$} & min\\\hline
        1 & 2 000 & 10 000 & 2 000 \\
        2 & 10 000 & 1 000 & 1 000\\
        3 & 40 000 & 1 000 & 1 000\\
        4 & 1 000 & 25 000 & 1 000\\\hline
        $\Sigma$ & 53 000 & 37 000 & 5 000\\
    \end{tabular}
\end{center}
Counting $\min(x,y)$ steps of the algorithm from \cite{SanTra07} for
intersecting lists of size $x$ and $y$ respectively, we get a total of 5\,000
steps for performing the intersection in all clusters whereas we get 37\,000
steps for intersecting the posting lists in an unclustered scenario -- more than
a factor five difference. More generally, for conjunctive queries where the
number of term occurrences in the clusters is not too highly correlated, we
expect improved performance from clustering. The main purpose of this paper is
to get a first idea how much this approach could help in practice.  After
introducing basic concepts in Section~\ref{s:prelim}, we formalize this idea in
Section~\ref{s:clustering} and develop an efficient clustering algorithm for the
resulting objective function.  This algorithm is based on the well known K-Means
principle but uses a fast multilevel scheme for initialization that may be of
independent interest.  In Section~\ref{s:experiments} we evaluate our approach
using large real world instances. Section~\ref{s:discussion} discusses the
results and possible future work.

\subsection*{More Related Work}

This paper is based on the diploma thesis of Jonathan Dimond \cite{Dimond13}.
Clustering has been previously proposed for accelerating full-text search, e.g.,
\cite[Section~7.1.6]{MRS08} following the \emph{clustering
  hypothesis}~\cite{van1979information} -- documents that are relevant to a
query tend to be more similar to each other than documents that are not
relevant. Documents similar to each other are clustered together.  Queries are
then executed via \emph{collection selection}.  Document clusters considered
irrelevant are disregarded and only clusters relevant to the query are used for
searching. Although this yields good scores in standard information retrieval
benchmarks, such an approach may lead to unexpected results in practice since
unsupervised learning techniques such as document clustering are notoriously
unreliable.

In many commercial applications one even wants -- at least as a
first step -- a complete and well defined set of results. For example, in SAP
HANA \cite{FCPBL12,TransierS10} the default mode of full-text search to is to
find \emph{all} documents matching a query. One reason is that the full text query is often only one of several filtering criteria in a complex SQL query. 
Further speedup techniques not based on document clustering have also been considered
including geographical tiering \cite{cambazoglu2009quantifying}, static index
pruning \cite{de2005improving} and dynamic index pruning
\cite{persin1994document}. However, all these techniques improve efficiency by
disregarding parts of the index.

\section{Preliminaries}
\label{s:prelim}

Let $D = \{d_1, \ldots, d_n\}$ denote the set of documents and
$N=\sum_{d\in D}|d|$ the total size of the corpus. Furthermore, let 
$T = \{t_1,\ldots, t_m\}$ denote the set of terms occurring in $D$.
Given a desired number of clusters $k$, we want to partition the documents into a
set of cluster $C = \{c_1,\ldots,c_k\}$ such that the expected query time is
small. 

\section{Clustering Based Search}
\label{s:clustering}

\subsection{Developing an Objective Function}
\label{ss:problem}

Of course, query costs depend on the query algorithm and the distribution of
queries. In order to come to an easy to handle objective function, we make some
assumptions and simplifications here that arguably lead to little loss in precision.

First of all, we focus on exact conjunctive queries involving exactly two terms.
Exact conjunctive queries with more than two terms are covered insofar as it is
usually a good idea to first intersect the lists for the two most rare terms and
often this takes most of the query time. We ignore ranking in this paper for
simplicity and because some applications require computing the full set of
answers, for example in a relational database context where further filtering
may throw away most of the results later.

In our implementation, we assume a two-level query algorithm that first
inspects an inverted index where each cluster is viewed as one document and
then inspects each cluster containing any document with the search terms. See \cite{Dimond13} for details. To simplify the exposition, in this section, we assume that the query cost is the sum of the per cluster query costs.
Let $\Phi(x,y)$ denote the cost for intersection two lists of length $x$ and $y$ respectively.
For the Lookup algorithm from \cite{SanTra07} we can approximate $\Phi(x,y)=\min(x,y)$.
For a particular query $(t,u)$ we then get cost 
$$\Psi_C(t,u)\Is\sum_{i=1}^k\Phi(n_i(t),n_i(u))$$
where $n_i(t)$ is the number of documents in cluster $i$ of clustering $C$ containing term $t$.
When the clustering $C$ is clear from the context, we omit the subscript `$C$'.

If we know the distribution of queries, we can now compute the expected cost of a query:
\begin{equation}
\psi\Is\expect[\Psi]\Is\sum_{\set{t,u}\in\binomial{T}{2}}\prob{(t,u)}\Psi(t,u)
\end{equation}
where $\prob{(t,u)}$ denotes the probability of observing query $(t,u)$.
Unfortunately, it is unrealistic to assume that we know the probability of all
queries so that we need an approximation. We thus only assume that we know the
probability of each query term. These probabilities can be estimated from a
query log or (less accurately) using statistics on the frequency of the terms in
the document collection itself.
If we now also assume that terms in a query
are chosen independently, we get
\begin{equation}\label{eq:psi}
\psi=\sum_{\set{t,u}\in\binomial{T}{2}}\prob{t}\prob{u}\Psi(u,c)\punkt
\end{equation}

\subsection{The Clustering Algorithm}
\label{ss:clustering}

Our starting point is an adaptation of the well known K-means algorithm to our
problem: In each iteration of the algorithm, each document $d$ is added to the
cluster where it ``fits'' best. In order to decide what the best fit is for
objective function $\psi$ from Equation~(\ref{eq:psi}) we only have to decide
how $\psi$ changes when $d$ is added. Hence, let $\score{j}{d}$ denote the
change in $\psi$ when document $d$ is added to cluster $j$. Similarly, let
$\score{j}{t}$ denote the change in $\psi$ when a document is added to cluster $j$
that contains only the single term $t$. We have $\score{j}{d}=\sum_{t\in
  d}\score{j}{t}$. Adding a single term $t$ to cluster $j$ only affects
a summand $\prob{t}\prob{u}\min(n_j(t),n_j(u))$ of $\phi$ if $n_j(t)<n_j(u)$. In
that case, $\min(n_j(t),n_j(u))$ increases by one. Hence,
$$\score{j}{t}=\prob{t}\sum_{u\neq t,n_j(t)<n_j(u)}\prob{u}\punkt$$
$\score{j}{t}$ can be computed in constant time using a lookup table that has to
be recomputed after each iteration.  Hence, $\score{j}{d}=\sum_{t\in
  d}\score{j}{t}$ can be computed in time linear in the document size.
Computing this lookup table itself can be done by sorting the terms occurring in
cluster $j$ by $n_j(t)$ and then computing a prefix sum over that
array. Assuming that the term frequencies are polynomial in the number of terms occurring in a cluster, sorting can
be done in linear time. Overall, one iteration of the K-means algorithm then
takes time (and space) $\Oh{kN}$.

\subsection*{Refinements}

The basic clustering algorithm defined above is already quite fast. However a
number of additional improvements are critical to scale to really large inputs.

\paragraph*{Multilevel Initialization.}
K-Means algorithms converge much faster if the initial solution is already of
high quality. We use a multilevel initialization that may be of independent
interest also in other applications. For a scaling factor $\scale<1$, we take a sample of size
$\max(k,\scale|D|)$ of the documents, cluster it recursively into $k$ clusters
(which is trivial for the base case $|D|=k$) and then run the K-means algorithm.
The K-means algorithm can lead to oscillations in the clustering that are
particularly pronounced when the clusters are small. When $D$ becomes small, we
therefore switch to an algorithm that updates the objective function after every
assignment of a document.

\paragraph*{TopDown Clustering.}
Since the running time of the K-Means algorithm grows at least linearly with the
number of clusters $k$, we use a hierarchical clustering algorithms that
recursively splits the documents: Subproblems with $s>|D|/k$ documents are
split into $\min(\chi,sk/|D|)$ pieces where the \emph{splitting factor} $\chi$ is a tuning parameter.
This way we obtain between $k$ and $2k$
clusters. An important side effect is that this approach
balances cluster sizes.

\paragraph*{Ignoring Infrequent Terms.}
Most search time is spent on queries involving long posting lists, i.e.,
frequent terms. Hence, the rare terms hardly contribute to the overall cost of
queries. Therefore, we can ignore rare terms while evaluating the
objective function without significantly affecting overall performance. 
This greatly accelerate the clustering algorithm and
simplifies its parallelization.

\paragraph*{Parallelization.}
Our implementation uses shared memory parallelization.  A massively parallel
distributed memory implementation is also easy as long as we can afford to store
word frequency statistics for each cluster and each frequent term on each node
of the system -- simply assign a subset of the documents to each node. In
connection with the above optimizations on TopDown clustering and ignoring
infrequent terms, this seems quite realistic even for huge document collections.

\subsection{Query Algorithms}
\label{ss:query}

As already explained, we focus on conjunctive queries involving two terms.  The
most direct way to use the clustering is to run the query on each cluster. When
the number of clusters $k$ is large, a possible improvement is to build a
\emph{cluster index} listing for each term which clusters contain documents with
this term. A query $(t,u)$ will then first intersect the list corresponding to
$t$ and $u$ in the cluster index. The query is then only forwarded to the
clusters containing both $t$ and $u$.  Note that a cluster index can be viewed
as an inverted index for a corpus with $k$ documents where each documents is the
concatenation of all the documents in a cluster.

We can also use the clustering to reorder the documents: The $j$-th document in
cluster $i$ gets document id $j+\sum_{\ell<i}|c_{\ell}|$.  Beyond this
reordering, the clustering is ignored -- we use the single-cluster Lookup
algorithm.  The motivation for this is the observation in \cite{SanTra07} that
nonuniform distribution of terms in the documents actually accelerates the set
intersection. Renumbering makes the distribution less uniform and thus may
accelerate the search.

\section{Implementation Details}
\label{s:implementation}

We have made a prototypical implementation using about 4\,000 lines of Haskell
and 2\,000 lines of C where all the time critical parts -- clustering and query
-- are implemented in C. We use OpenMP for parallelization\footnote{Source are
  here: \url{https://www.github.com/jdimond/diplomarbeit}}. We switch to
document grained updates of the objective function once $|D|<100k$.  
The default shrink factor for the multilevel initialization is $\scale=0.1$.
The K-means algorithm repeats as long as the objective function value improves by at least 1~\%.
The splitting factor for hierarchical clustering is set to $\chi=8$.
Only the $\mathrm{TC}=100\,000$ most frequent terms are used for clustering.
The Lookup algorithm \cite{SanTra07} uses bucket size $16$ (8 for the cluster index). 
This seems a good tradeoff between space and speed requirements.

\section{Experiments}
\label{s:experiments}

\begin{table}[htb]
    \centering
    \begin{tabular}{lrrrr}
                  & GOV2           & GOV2s & Wikipedia & pagenstecher.de \\
        \hline
        Documents & \num{25205179}  & \num{631975969} & \num{6096279} & \num{786474} \\
        Terms     & \num{38562580} & \num{25221691} & \num{12295297} & \num{573725} \\
        Terms / Document & \num{652.22} & \num{18.19} & \num{230.54} & \num{35.70} \\
        \hline
        Input size (raw) & \SI{396.74}{GB} & \SI{396.74}{GB} & \SI{12.37}{GB} & \SI{175.14}{MB} \\
        Inverted Index size & \SI{16.25}{GB} & \SI{32.83}{GB} & \SI{3.09}{GB} & \SI{89.8}{MB} \\
    \end{tabular}\vspace*{-2mm}
    \caption{Dataset Statistics}
    \label{tab:datasetstatistics}
\end{table}

\begin{table}[htb]
    \centering
    \begin{tabular}{lrrrr}
                  & AOL           & Wikipedia & pagenstecher.de \\\hline
        Queries & \num{29077553}  & \num{11000000} & \num{13230} \\
        Distinct Terms & \num{1501946} & \num{1067091} & \num{981} \\
    \end{tabular}\vspace*{-2mm}
    \caption{Query Log Statistics}
    \label{tab:querylogs}
\end{table}

\begin{figure}[htb]
    \centering
    \input{querystats.tex}\vspace*{-7mm}
    \caption{Probability of a term appearing in a query as a function of its
    rank on a log-log scale. A sample of 100 terms with exponentially growing
ranks is plotted.}
    \label{fig:querystats}
\end{figure}

\begin{figure}
    \centering
    \input{sizes.tex}
    \caption{Speedups for different number of clusters (flat clustering).}
    \label{fig:speedup_sizes}
\end{figure}
\begin{figure}
    \centering
    \input{gov2s_sizes.tex}
    \caption{Speedups for different number of clusters (TopDown clustering).}
    \label{fig:gov2s_sizes}
\end{figure}

\begin{figure}
    \centering
    \input{doclimits.tex}
    \caption{Speedups for GOV2 with varying number of documents $|D|$ for
        $k=\num{2500}$ clusters.}
    \label{fig:doclimits}
\end{figure}

\begin{figure}[htb]
    \centering
    \input{tdvsfm.tex}\vspace*{-7mm}
    \caption{Speedups comparison for flat (FM) and TopDown (TD) clustering (GOV2).}
    \label{fig:td_vs_fm}
\end{figure}

\begin{figure}[htb]
    \centering
    \input{times_numclusters_fm.tex}
    \input{times_numclusters_td.tex}
    \caption{Clustering times $[s]$ for flat (left) and TopDown (right) clustering for 10 independent trials.
    \label{fig:times_numclusters}}
\end{figure}
\begin{figure}[htb]
    \centering
    \input{numterms.tex}
    \caption{Speedups for GOV2 with varying $TC$ for $K=2\,500$.}
    \label{fig:numterms}
\end{figure}

All experiments were done on a machine with two octa-core Intel Sandy Bridge
Xeon E5-2670 processors with 2.6GHz and 64 GB RAM (i.e., 16 cores and 32
hardware threads). The operating system was SuSE Linux Enterprise Server 11
(kernel version 3.0.42). The compilers used were GHC 7.6.2 and GCC 4.7.2 with
optimization level \texttt{-O3}. Clustering is run in parallel. Queries are run sequentially.

Table~\ref{tab:datasetstatistics} gives the text corpora used for our
benchmarks.  GOV2 \cite{clarke2004gov2} is one of the standard benchmarks used
in the literature.  GOV2s is the same corpus but each \emph{sentence} is used as
one document. The sentences were extracted using the Stanford NLP library \cite{MSBFBM14}. This emulates a corpus with many very small documents as you may
find it in corporate databases and a situation where you are looking for terms
occurring together in the same sentence. Wikipedia is the plain text contained
in the articles in the English Wikipedia in May 3, 2013. Pagenstecher.de
contains user posts from a German online community for car tuning. This corpus
was selected because we have an authentic query log for it.

Further query logs used are shown in Table~\ref{tab:querylogs}.  The AOL log contains
the two term queries from \cite{pass2006picture}.  We use this log as
semi-realistic input for our GOV2 and GOV2s test collections.  For the Wikipedia
corpus, we generate a synthetic log: for each article reference, we add all pairs of
terms in the title of the referenced article to the log. Figure~\ref{fig:querystats} shows the distribution of term frequencies in
these logs. We can see that all of them, including the synthetic Wikipedia log,
show a Zipf-like distribution of term frequencies.

Our main concern are speedups over the single cluster case. We distinguish the
``theoretical'' speedup $S_T$ predicted by the function $\psi$ from
Equation~(\ref{eq:psi}), the speedup $S_C$ obtained using the cluster index from
Section~\ref{ss:query}, and the speedup $S_R$ using the reordering algorithm
from Section~\ref{ss:query}.  Figures~\ref{fig:speedup_sizes} and
Figures~\ref{fig:gov2s_sizes} compare these values for varying number $k$ of
clusters. Most of the time, $S_T$ overestimates the speedup but it seems
strongly correlated with the actual behavior -- which is all we need for making
it a useful objective function for the clustering algorithm. Generally,
increasing the number of clusters helps to increase speedup. However, for the
Wikipedia instance using the cluster index, the speedup decreases for $k>500$ --
it is clear that eventually, overheads for the additional indirection start to
show. A surprise is that the reordering algorithm achieves much better
performance than the cluster index. Overall, we achieve speedups between 1.3 and
4 which is not overwhelming but certainly significant and possibly useful.  The
Pagenstecher instance seems very different from the others -- it achieves much
higher speedups and it is the only one where the theoretical speedup is smaller
than the practical speedups. This may simply be due to its relatively small size
or specialized topic but it is also the only one with truly realistic query
logs. If it turns out that this is the reason for the performance difference, we
might hope for larger speedups also for large
instances. Figure~\ref{fig:doclimits} gives some reason for optimism here since
it indicates that speedups may actually \emph{increase} with growing number of
documents. 

Figure~\ref{fig:td_vs_fm} indicates that the flat clustering algorithm gives
slightly better speedups than the TopDown algorithm. However, for the values of
$k$ that give good speedup, we do not really have a choice --
Figure~\ref{fig:times_numclusters} shows that the running time of the flat
clustering algorithms actually grows \emph{superlinearly} with the number of
clusters $k$ -- apparently, the algorithm also converges more slowly for large
$k$. The TopDown algorithm is orders of magnitude faster than the flat
algorithm. Indeed, our clustering algorithm needs much less time than the time
for parsing and indexing the documents. Hence, the preprocessing overhead is not a
big issue. 

Usually, preprocessing techniques also come with a penalty for storing the
preprocessed information. However, in our case the contrary is the case here --
clustering allows better compression of the posting lists, see
Appendix~\ref{app:compression} for details.

It might be argued that it is risky that our objective function is tied to a
particular intersection algorithm. To assess this risk we have evaluated the
theoretical speedup for a different cost function that assumes an asymptotically
optimal comparison based intersection algorithm with running time
$\Phi(x,y)=x\log\frac{y}{x}$ for $x>y$ (e.g., \cite{BaezaYates2004}). Appendix~\ref{app:comparison} indicates
that this gives very similar results -- indeed, the theoretical speedups are
even higher than for the lookup algorithm even though $\min(x,y)$ was used as
the objective function for clustering.

In Figure~\ref{fig:numterms} we investigate how the number of terms (TC) we use
for clustering influences speedup. All three speedup measure show that even the
10\,000 most frequent terms would be enough for the GOV2 input. This is good
news since this allows faster clustering algorithm. In particular, a massively
parallel clustering algorithm can probably afford to replicate all lookup tables
over all processors.

\section{Conclusions}
\label{s:discussion}

We have demonstrated that document clustering can significantly accelerate
conjunctive queries while still giving exact results. Using a multilevel
hierarchical clustering algorithm we were able to do high quality clustering
even faster than the needed for parsing and indexing the documents.
Our approach of tailoring the objective function for clustering to the actual 
performance the query algorithm might also be useful in other situations like clustering for inexact search or in order to compress data.

Several ideas suggest themselves for further improving the results.  Since the
approach is most useful for big inputs, scaling to even bigger corpora may be the
main concern. This seems feasible since a massively parallel implementation is
relatively easy. We have not done so yet since it naturally requires a lot of
resources.

Equally interesting are efforts to increase the quality of the clustering.
Currently, we only look at the impact of adding documents to a cluster. However,
for small clusters (e.g., during initialization) it also matters how removing a
document from a cluster affects query performance. We believe that our
lookup-table approach can take this into account without big performance
penalties. Also, compromises between the pure round based K-means algorithm and
the variant with document-wise updates seem possible which might improve
convergence speed and overall quality. Also further tuning of the TopDown
algorithm seems promising. Probably one wants to use larger splitting factors at
least at the top of the recursion tree in order to improve quality.

Since using the clustering for reordering posting lists was very successful,
this should also be considered more closely. In particular, we would like to
have a more symmetric version of the lookup algorithm for list
intersection. Currently, the algorithm traverses the shorter list while making
lookups in the longer list. Actually, we would like to have a more adaptive
algorithm that scans the list which is more dense at the current position.  For
example, when a lookup finds an empty bucket, we might switch to the other list.
Also, it is quite clear that for clustering for reordering should (recursively)
go all the way down to clusters consisting only of a few documents. To make this
efficient, we need to dynamize our cutoffs -- we only want to consider terms
frequent in the remaining documents.

Last but not least, we have to investigate what happens for other types of
queries.  Perhaps most interesting are top-K queries where we are only
interested in the most relevant results using some scoring function. Once more,
we want to exploit the information inherent in the clustering to gain
performance but the clustering must not influence the result. The hope might be
that a cluster $c$ which contains only few documents with a certain query term then $c$
might contain even less \emph{relevant} documents with that term so that only a
fraction of its posting list actually needs to be considered.

\bibliographystyle{abbrv}
\bibliography{diss,thesis/bibliography}

\begin{appendix}
\section{Data Compression}\label{app:compression}
 
Figure~\ref{fig:compression} shows the space consumption per posting list entry
using several encoding techniques. An interesting observation is that Golomb
coding \cite{golomb1966runlength} is best without clustering whereas Elias-$\gamma/\delta$ coding \cite{elias1975universal,baeza1999modern} is better
with clustering. The reason is that these encoding schemes can better adapt to
varying distances. Considering that data compression was not our primary
objective, the savings are considerable.
\begin{figure}[htb]
    \centering
    \input{compression.tex}
    \caption{Compression of the inverted index using different clustering
    algorithms and encodings on the GOV2 dataset with $k=\num{1280}$
    clusters.}
    \label{fig:compression}
\end{figure}
\section{Comparison Based Intersection}\label{app:comparison}

Figure~\ref{fig:speedups_log} compares the theoretical speedup obtained using the cost function $\Phi(x,y)=\min(x,y)$ ($S_T$) and using the function stemming from comparison based list intersection \cite{BaezaYates2004} ($S_L$). We see that for a comparison based list intersection algorithm, even higher speedups are predicted.
\begin{figure}[htb]
    \centering
    \input{log.tex}
    \caption{Speedups $S_T$ and $S_L$ on different datasets using the adapted
    cost function for $S_L$. The clusterings have $K=\num{1280}$ clusters and
    use the FMClustering algorithm with exception of GOV2-TD.}
    \label{fig:speedups_log}
\end{figure}

\end{appendix}
\end{document}